\documentstyle[aps,12pt]{revtex}

\begin{document}
\author{E.N. Glass$^{\dagger }$ and Jonathan Kress$^{\ddagger }$}
\address{$^{\dagger }$Physics Department, University of Michigan,\\
Ann Arbor, MI 48109, USA\thanks{%
Permanent Address: Physics Department, University of Windsor, Ontario, Canada%
}\\
$^{\ddagger }$School of Mathematics and Statistics, \\
University of Sydney, NSW 2006, Australia }
\date{September 27, 1998}
\title{Solutions of Penrose's Equation}
\maketitle

\begin{abstract}
\\\\The computational use of Killing potentials which satisfy Penrose's
equation is discussed. Penrose's equation is presented as a conformal
Killing-Yano equation and the class of possible solutions is analyzed. It is
shown that solutions exist in spacetimes of Petrov type $O$, $D$ or $N$. In
the particular case of the Kerr background, it is shown that there can be no
Killing potential for the axial Killing vector. \newpage\ 
\end{abstract}

\section{INTRODUCTION}

In a spacetime which admits a Killing vector $k^a$ it is straightforward to
find its Killing potential. Killing potentials are real bivectors $Q^{ab}$
whose divergence returns the Killing vector, $(1/3)\nabla _bQ^{ab}=k^a$.
Killing potentials attain physical importance when they are used in the
Penrose-Goldberg (PG) \cite{josh} superpotential for computing conserved
quantities such as mass and angular momentum. The PG superpotential is 
\begin{equation}
U_{PG}^{ab}=\sqrt{-g}\frac 12G^{ab}{}_{cd}Q^{cd},  \label{pgpot}
\end{equation}
where $G^{ab}{}_{cd}=-^{*}R^{*ab}{}_{cd},$ the negative right and left dual
of the Riemann tensor. When the Ricci tensor is zero then $%
G^{ab}{}_{cd}=C^{ab}{}_{cd}$, the Weyl tensor. If $Q^{ab}$ satisfies
Penrose's equation (\ref{pgq}) then 
\begin{equation}
\nabla _bU_{PG}^{ab}=\sqrt{-g}G^{ab}k_b  \label{divpgpot}
\end{equation}
for Einstein tensor $G^{ab}$. The current density 
\begin{equation}
J^a=\sqrt{-g}G^{ab}k_b  \label{4current}
\end{equation}
is conserved independently of the left side of Eq.(\ref{divpgpot}). It is
the PG superpotential that allows the Noether quantities to be computed by
integrating over closed two-surfaces, which is Penrose's quasi-local
construction \cite{roger}. If one views the Killing vector itself as a
conserved current then its integral over a three-surface is identically
equal to $1/3$ the integral of its Killing potential over the bounding
two-surface and no new information can be obtained.

The tensor version of Penrose's equation \cite{pen&rin} is 
\begin{equation}
{\sc P}^{abc}:=\nabla^{(a}Q^{b)c}-\nabla^{(a}Q^{c)b}
+g^{a[b}Q^{c]e}{}_{;e}=0.  \label{pgq}
\end{equation}
With $j^a:=(1/3)\nabla _bQ^{*ab}$, and $k^a:=(1/3)\nabla _bQ^{ab}$, an
equivalent equation \cite{tod} to ${\sc P}^{abc}=0$ is 
\begin{equation}
\nabla _cQ^{ab}=-2\delta _c^{[a}k^{b]}+2(\delta _c^{[a}j^{b]})^{*}.
\label{altpeqn}
\end{equation}
If $Q^{ab}$ is a solution of the Penrose equation then $%
k_{(b;c)}=-(1/2)Q_{a(b}R^a{}_{c)}$ with a similar relation connecting $j^a$
and $Q^{*ab}$. For Ricci-flat spacetimes $j^a$ and $k^a$ are Killing vectors.

For a particular spacetime the number of independent Killing vectors is
between zero and ten. Penrose \cite{pen&rin} gave the complete solution to
Eq.(\ref{pgq}) in Minkowski space for ten real independent $Q^{ab}$.

This work discusses the existence of Killing potentials which satisfy
Penrose's equation or equivalently the conformal Killing-Yano (CKY) equation
for 2-form ${\cal Q}$. The fact that such tensors only exist in spacetimes
of Petrov type $D$, $N$ or $O$ is discussed in section \ref{CKYexist} and
Appendices \ref{App:shearfree} and \ref{App:CKYint}.

In the Kerr background, it has previously been shown that there is no
Killing potential for the axial Killing vector \cite{englass}. We show, in
section \ref{divCKY}, how this can be anticipated from properties of the
curvature and the fact that the axial Killing vector must vanish along the
axis of symmetry.

We use both the abstract index notation familiar to relativists and some
coordinate free notation for which we provide Appendix \ref{App:forms} as a
reference. We use bold face for index free tensor notation excepting
differential forms which appear in calligraphic type. Appendix \ref
{App:Petrov} describes some aspects of the Petrov classification in a way
convenient for our purposes.

\section{PREVIOUS RESULTS}

An exact solution of the Penrose equation for Kerr's vacuum solution is
given below in Eq.(\ref{qt}). This solution was first used in the context of
the PG superpotential construction in \cite{ed-mark}. The Kerr solution has
two Killing vectors (KVs), stationary $k_{(t)}$ and axial $k_{(\varphi )}$,
and the metric is 
\begin{equation}
g^{Kerr}=l\otimes n+n\otimes l-m\otimes \bar{m}-\bar{m}\otimes m\,,
\label{kerr}
\end{equation}
where $\{l,n,m,\bar{m}\}$ is the Newman-Penrose principal null coframe,
given in Boyer-Lindquist coordinates by 
\begin{eqnarray}
l &=&dt-(\Sigma /\Delta )\,dr-a\sin ^2\theta \,d\varphi \,,  \label{npkerr}
\\
n &=&\frac \Delta {{2\Sigma }}[dt+(\Sigma /\Delta )\,dr-a\sin ^2\theta
\,d\varphi ]\,,  \nonumber \\
m &=&\frac 1{\sqrt{2}\bar{R}}[\text{i}a\text{sin}\theta \,dt-\Sigma
\,d\theta -\text{i(}r^2+a^2)\text{sin}\theta \,d\varphi ]\,,  \nonumber
\end{eqnarray}
where $R=r-ia\cos \theta $, $\Sigma =R\bar{R}$ and $\Delta =r^2+a^2-2m_0r$.
The Killing potential for $k_{(t)}$ is the bivector $Q_{(t)}^{ab}$ obtained
by raising the components of the 2-form 
\begin{equation}
{\cal Q}_{(t)}=-\left( R{\cal M}+\bar{R}\bar{{\cal M}}\right) \,,  \label{qt}
\end{equation}
where ${\cal M}:=l\wedge n-m\wedge \bar{m}$ \ is an anti self-dual 2-form,
that is $*{\cal M}=-i{\cal M}$. We mention that $Q_{(t)}^{ab}$ is a global
solution since the quasi-local PG mass, resulting from integration of the PG
superpotential over two-surfaces of constant $t$ and $r$, is {\em %
independent of choice of two-surface} 
\begin{equation}
\displaystyle \oint 
\limits_{S^2}U_{PG}^{ab}\ dS_{ab}=-8\pi \ m_0  \label{pgmass}
\end{equation}
for any $r$ beyond the outer event horizon.

The next interesting result involves the axial Kerr symmetry. Goldberg \cite
{josh} found asymptotic solutions of the Penrose equation for the
Bondi-Sachs metric which includes the Kerr solution as a special case. But
Glass \cite{englass} showed that the axial Killing potential could not be a
solution of the Penrose equation at finite $r$.

The bivector $Q_{(t)}^{ab}$ generally has six independent components and so
enough information to describe two Killing vectors. Since the Kerr solution
has two KVs, can the dual of $Q_{(t)}^{ab}$ yield $k_{(\varphi )}$? Direct
differentiation shows 
\begin{equation}
\nabla _bQ_{(t)}^{*ab}=0,  \label{divdual}
\end{equation}
and so $Q_{(t)}^{ab}$ can only yield $k_{(t)}$. In fact $Q_{(t)}^{*ab}$
satisfies the Killing-Yano (KY) equation, which for an antisymmetric tensor $%
A_{ab}$ can be written as 
\begin{equation}
A_{a(b;c)}=0.  \label{kyeqn}
\end{equation}
This generalizes Killing's equation to antisymmetric tensors and can be
further generalized to antisymmetric tensors of arbitrary valence. Modern
usage reserves the name KY tensor for antisymmetric tensors. For the Kerr
solution a symmetric tensor $K_{ab}$ is constructed from the dual Killing
potential by 
\begin{equation}
K_{ab}=Q_{(t)a}^{*e}Q_{eb}^{*(t)}=2\Sigma \ l_{(a}n_{b)}-r^2g_{ab}.
\label{killpot}
\end{equation}
This ``hidden'' symmetry of the Kerr solution was discovered Carter \cite
{carter} and later shown to be the ``square'' of a two-index Killing spinor 
\cite{walkpen}, or equivalently, the ``square'' of a Killing-Yano tensor.
Though $K_{ab}$ satisfies Eq.(\ref{kyeqn}) it is symmetric and generally
referred to as a Killing tensor.

Collinson \cite{col} found that all vacuum metrics of Petrov type $D$, with
the exception of Kinnersley's type $IIIB$, possess a KY tensor. He gave an
explicit expression for both the KY tensor and it's associated Killing
tensor.

\section{EXISTENCE OF SOLUTIONS}

\subsection{Conformal Killing-Yano Tensors}

Many of the arguments in this work depend on the conformal covariance of
Penrose's equation. Penrose and Rindler\cite{pr2} established the conformal
covariance of its spinor form $\nabla _{A^{\prime }}{}^{(A}\sigma ^{BC)}=0$
for a symmetric spinor $\sigma ^{BC}$. The tensor version was previously
discovered by Tachibana as the conformally covariant generalization of the
KY equation \cite{tach}. In this paper it was written in the form 
\begin{equation}
Q_{a(b;c)}={(1/3)}\left[
g_{bc}Q_a{}^e{}_{;e}-g_{a(b}Q_{c)}{}^e{}_{;e}\right] \,.  \label{pencky}
\end{equation}
In that same work Tachibana showed that in a Ricci-flat space, for $Q_{ab}$
a CKY bivector satisfying Eq.(\ref{pencky}), $(1/3)\nabla ^bQ_{ab}$ is a
Killing vector.

From Eq.(\ref{pencky}) we can obtain an expression for $Q_{ab;c}$ by writing
out the symmetrization brackets explicitly: 
\[
Q_{ab;c}=-Q_{ac;b}+{\frac 23}g_{bc}Q_a{}^e{}_{;e}-{\frac 13}%
g_{ab}Q_c{}^e{}_{;e}-{\frac 13}g_{ac}Q_b{}^e{}_{;e}\,.
\]
Now, since $Q_{ab;c}$ is antisymmetric in the first two indices, we have 
\begin{eqnarray*}
3Q_{ab;c} &=&Q_{ab;c}+Q_{ab;c}-Q_{ba;c} \\
&=&Q_{ab;c}-Q_{ac;b}+{\frac 23}g_{bc}Q_a{}^e{}_{;e}-{\frac 13}%
g_{ab}Q_c{}^e{}_{;e}-{\frac 13}g_{ac}Q_b{}^e{}_{;e} \\
&&\ \ \qquad +Q_{bc;a}-{\frac 23}g_{ac}Q_b{}^e{}_{;e}+{\frac 13}%
g_{ba}Q_c{}^e{}_{;e}+{\frac 13}g_{bc}Q_a{}^e{}_{;e}
\end{eqnarray*}
and so from (\ref{pencky}) we can deduce that 
\begin{equation}
3Q_{ab;c}=3Q_{[ab;c]}-2g_{c[a}Q_{b]}{}^e{}_{;e}\,.  \label{cky2}
\end{equation}
It is easily verified that given Eq.(\ref{cky2}) we recover Eq.(\ref{pencky}%
) and hence Eq.(\ref{cky2}) is an alternative form of the CKY equation.
Furthermore Penrose's Eq.(\ref{pgq}) can easily be rewritten as Tachibana's
Eq.(\ref{pencky}) and so is another form of the CKY equation.

Since $Q$ is an antisymmetric tensor, it is natural to discuss it's
properties in the language of differential forms. Equation (\ref{cky2}) is
manifestly antisymmetric in the first two indices, and so it is
straightforward to verify that it is the abstract index equivalent of the
CKY 2-form equation of Benn {\it et al} \cite{b&c&k}, 
\begin{equation}
3\nabla _Z{\cal Q=}Z%
\hbox{\vrule  height 0ex     width 1ex      depth 0.11ex
                 \vrule  height 1.6ex   width 0.11ex   depth 0.11ex
                 \vrule  height 0ex     width 0.5ex    depth 0ex   }d{\cal Q}%
-Z^{\flat }\wedge \delta {\cal Q},\qquad \forall Z\,.  \label{formcky2}
\end{equation}
In this form, since $*$ commutes with $\nabla _Z$, it is readily verified
using the identities given in Appendix \ref{App:forms}, that whenever ${\cal %
Q}$ is a CKY 2-form so is $*{\cal Q}$. Thus any solution to the CKY equation
can be decomposed into self-dual and anti self-dual CKY 2-forms.

\subsection{Existence of CKY 2-forms}

\label{CKYexist}On a flat background the CKY equation has many solutions,
while, as will be explained, in a more general spacetime the curvature
imposes tight consistency conditions and there can be at most two
independent solutions, one self-dual and one anti self-dual with respect to
the Hodge star. This result appears to be closely tied to the
four-dimensional nature of spacetime and the properties of these solutions
are almost universally discussed in their spinor form, where the utility of
the two-component spinor formalism simplifies the calculations. A detailed
discussion of this can be found in spinor form in \cite{b&c&k} or in terms
of differential forms in \cite{kress}.

Since any CKY 2-form can be decomposed into self-dual and anti self-dual
parts that are themselves CKY 2-forms, in discussing their existence, it is
sufficient to consider only 2-forms of definite Hodge-duality.

In order to understand how the curvature of the underlying spacetime
restricts the solutions to Eq.(\ref{formcky2}) two steps are required.
Firstly, it can be shown directly from the CKY 2-form equation that the real
eigenvectors of (anti) self-dual CKY 2-forms are shear-free and hence
principal null directions of the conformal tensor. Secondly, by
differentiating Eq.(\ref{formcky2}) an integrability condition can be
obtained that restricts the Petrov type by showing these eigenvectors to be 
{\em repeated} principal null directions.

In the case of non-null self-dual 2-forms, Dietz and R\"{u}diger \cite{d&r}
used spinor methods to obtain both of these results for a scaling covariant
generalization of Eq.(\ref{formcky2}). It was later shown, again using
spinor methods, that similar results can be obtained for the null case \cite
{b&c&k}.

An outline of these results in the notation of differential forms is given
in Appendices \ref{App:shearfree} and \ref{App:CKYint}. It is shown that
apart from conformally flat spacetimes, non-null (anti) self-dual CKY
2-forms can only exist in spacetimes of Petrov type $D$, while null (anti)
self-dual CKY 2-forms require a background spacetime of Petrov type $N$.

\subsection{The divergence of a CKY 2-form}

\label{divCKY}In order to apply the PG superpotential method using a given
CKY 2-form ${\cal Q}$, its divergence (coderivative) $\delta {\cal Q}$ must
be dual to a Killing vector. Tachibana showed that this was always the case
in a Ricci flat background \cite{tach} (the result also holds for the
slightly more general case of an Einstein spacetime).

In the Kerr background, there are two independent Killing vectors and two
independent CKY 2-forms (one of each Hodge-duality). However the divergence
of either of these CKY 2-forms is proportional to the timelike Killing
vector, leaving the axial KV without a Killing potential. This allows a
divergence free linear combination of the self-dual and anti self-dual CKY
2-forms to be found. The Hodge-dual of this 2-form is known as a
Killing-Yano 2-form and satisfies the Killing-Yano equation (\ref{kyeqn}),
which can be written in a similar fashion to Eq.(\ref{formcky2}) as 
\begin{equation}
3\nabla _X{\cal Q}=X%
\hbox{\vrule  height 0ex     width 1ex      depth 0.11ex
                 \vrule  height 1.6ex   width 0.11ex   depth 0.11ex
                 \vrule  height 0ex     width 0.5ex    depth 0ex   }d{\cal Q}%
\,.  \label{ky2form}
\end{equation}
However, this leaves open the question of why it is the timelike rather than
the axial KV that possesses a Killing potential? To answer this question, we
note that the axial Killing vector must vanish along the symmetry axis and
we show that a Killing vector obtained as the divergence of a CKY 2-form
must be nowhere vanishing.

First consider a non-null anti self-dual CKY 2-form ${\cal Q}^{-}$. From Eq.(%
\ref{formcky2}) we can write $d({{\cal Q}^-}^2)$ in terms of ${\cal Q}^{-}$
and $\delta {\cal Q}^{-}$: 
\[
d\left( {{\cal Q}^-}^2\right) =\mbox{$4\over3$} \left( \delta {\cal Q}%
^{-}\right) ^{\sharp }%
\hbox{\vrule  height 0ex     width 1ex      depth 0.11ex
                 \vrule  height 1.6ex   width 0.11ex   depth 0.11ex
                 \vrule  height 0ex     width 0.5ex    depth 0ex   } {\cal Q}%
^{-}\,, 
\]
which after contracting with ${\cal Q}^{-}$ leads to 
\[
\delta {\cal Q}^{-}=-\mbox{$3\over2$} \left( d({{\cal Q}^-}^2)\right)
^{\sharp }%
\hbox{\vrule  height 0ex     width 1ex      depth 0.11ex
                 \vrule  height 1.6ex   width 0.11ex   depth 0.11ex
                 \vrule  height 0ex     width 0.5ex    depth 0ex   } {\cal Q}%
^{-}\,. 
\]
Hence $\delta{\cal Q}^{-}$ vanishes if and only if $d({{\cal Q}^-}^2)$
vanishes.

In a vacuum type $D$ background we can deduce that ${{\cal Q}^-}^2$ is a
constant multiple of ${\Psi _2}^{-{\frac 23}}$ from the fact that ${\cal Q}%
^{-}$ is an eigen-2-form of ${\bf C}$ and both $\left( {{\cal Q}^-}%
^2\right)^{-{\frac 32}}{\cal Q}^{-}$ and ${\bf C}{\cal Q}^-$ are Maxwell
fields. Hence if ${\cal Q}^-$ vanishes, then so does $\Psi_2$ and the
background becomes conformally flat.

Further, it can be deduced from the Bianchi identities that for a type $D$
vacuum spacetime, the gradient of $\Psi _2$ vanishes if and only if the $%
\Psi _2$ itself vanishes. (In the Newman-Penrose formalism, using a
principal null tetrad, the vacuum type $D$ condition implies that the only
nonzero curvature component is $\Psi _2$ and $\kappa =\sigma =\nu =\lambda
=0 $. Then, imposing $\nabla_{X_a}\Psi _2=0$, the Bianchi identities lead to
either $\rho =\mu =\tau =\pi =0$ or $\Psi _2=0$. If we assume the former,
then the NP equations for the derivatives of the spin coefficients
immediately force the conclusion that $\Psi _2$ vanishes.) We therefore
conclude that ${{\cal Q}^-}^2$ is nowhere constant and hence $\delta {\cal Q}%
^-$ is nowhere vanishing and Kerr's axial Killing vector {\em cannot} have a
Killing potential.

\section{SUMMARY}

We have shown here that Penrose's equation for Killing potentials is
equivalent to the conformal Killing-Yano equation for 2-forms. With no
appeal to Ricci-flatness, existence of solutions was proven for spacetimes
of Petrov type $D$, $N$ or $O$. It was further shown, for type $D$ vacuum
backgrounds possessing a Killing-Yano 2-form, that Killing vectors with
zeros cannot have Killing potentials.

\begin{center}
{\bf ACKNOWLEDGMENTS}
\end{center}

E.N. Glass was partially supported by an NSERC of Canada grant and thanks
the Physics Department of the University of Michigan and Professor Jean
Krisch for their hospitality. Jonathan Kress began this work at the
University of Newcastle.

\appendix 

\section{DIFFERENTIAL FORMS}

\label{App:forms}

We denote a basis for vector fields by $\{X_a\}$. The natural dual of this
we denote by $\{e^a\}$, a basis for covector or 1-form fields. A coordinate
basis is $X_a={\frac{\partial}{{\partial x^a}}}$ and $e^a=dx^a$. The metric
gives a natural bijection between vector and 1-form fields, which we denote
by ${}^\sharp$ and ${}^\flat$; $X^\flat$ is the 1-form metric dual to the
vector $X$ and $\alpha^\sharp$ is the vector field metric dual to the 1-form 
$\alpha$.

The 1-forms, along with the wedge product $\wedge $, generate the algebra of
differential forms. The wedge product is anti-symmetric and so the
differential forms of degree $p$ can be thought of as the subset of
covariant tensors of valence $p$ that are antisymmetric in their arguments.
If $\alpha $ and $\beta $ are 1-forms with components $\alpha _a=\alpha
(X_a) $ and $\beta _a=\beta (X_a)$, then 
\begin{equation}
\alpha \wedge \beta =\alpha _{[a}\beta _{a]}e^a\otimes e^b=\alpha _a\beta
_be^a\wedge e^b\,.  \label{wedge}
\end{equation}

A vector can be contracted with the $p$-form ${\cal P}$ to give a $(p-1)$%
-form $X%
\hbox{\vrule  height 0ex     width 1ex      depth 0.11ex
                 \vrule  height 1.6ex   width 0.11ex   depth 0.11ex
                 \vrule  height 0ex     width 0.5ex    depth 0ex   } {\cal P}
$ so that 
\[
(X%
\hbox{\vrule  height 0ex     width 1ex      depth 0.11ex
                 \vrule  height 1.6ex   width 0.11ex   depth 0.11ex
                 \vrule  height 0ex     width 0.5ex    depth 0ex   } {\cal P}%
)(X_{a_1},X_{a_2},\ldots ,X_{a_{p-1}}) =p{\cal P}(X,X_{a_1},X_{a_2},\ldots
,X_{a_{p-1}})\,, 
\]
and so the components of a $p$-form can be expressed using the hook as 
\[
{\cal P}_{ab\ldots c}={\cal P}(X_a,X_b,\ldots ,X_c) ={\frac 1{{p!}}}X_c%
\hbox{\vrule  height 0ex     width 1ex      depth 0.11ex
                 \vrule  height 1.6ex   width 0.11ex   depth 0.11ex
                 \vrule  height 0ex     width 0.5ex    depth 0ex   } \ldots%
\hbox{\vrule  height 0ex     width 1ex      depth 0.11ex
                 \vrule  height 1.6ex   width 0.11ex   depth 0.11ex
                 \vrule  height 0ex     width 0.5ex    depth 0ex   } X_b%
\hbox{\vrule  height 0ex     width 1ex      depth 0.11ex
                 \vrule  height 1.6ex   width 0.11ex   depth 0.11ex
                 \vrule  height 0ex     width 0.5ex    depth 0ex   } X_a%
\hbox{\vrule  height 0ex     width 1ex      depth 0.11ex
                 \vrule  height 1.6ex   width 0.11ex   depth 0.11ex
                 \vrule  height 0ex     width 0.5ex    depth 0ex   } {\cal P}%
\,. 
\]
We can define an inner product between any pair of 2-forms: 
\[
{\cal P}\cdot {\cal Q}={\frac 12}X_a%
\hbox{\vrule  height 0ex     width 1ex      depth 0.11ex
                 \vrule  height 1.6ex   width 0.11ex   depth 0.11ex
                 \vrule  height 0ex     width 0.5ex    depth 0ex   } X_b%
\hbox{\vrule  height 0ex     width 1ex      depth 0.11ex
                 \vrule  height 1.6ex   width 0.11ex   depth 0.11ex
                 \vrule  height 0ex     width 0.5ex    depth 0ex   } {\cal P}%
\ X^a%
\hbox{\vrule  height 0ex     width 1ex      depth 0.11ex
                 \vrule  height 1.6ex   width 0.11ex   depth 0.11ex
                 \vrule  height 0ex     width 0.5ex    depth 0ex   } X^b%
\hbox{\vrule  height 0ex     width 1ex      depth 0.11ex
                 \vrule  height 1.6ex   width 0.11ex   depth 0.11ex
                 \vrule  height 0ex     width 0.5ex    depth 0ex   } {\cal Q}
=2{\cal P}_{ab}{\cal Q}^{ab}\,. 
\]
For ${\cal P}\cdot{\cal P}$ we write ${\cal P}^2$.

The metric defines a natural map from $p$-forms to $(n-p)$-forms called the
Hodge star. In four dimensions, this maps 2-forms to 2-forms, and is defined
so that 
\[
{\cal P}\wedge *{\cal Q}=\left( {\cal P}\cdot {\cal Q}\right) *1\,, 
\]
where $*1$ is the volume 4-form. For a Lorentzian metric, this map squares
to $-1$ and so has eigenvalues, $\pm i$. Elements of the eigenspace
corresponding to $(-i)+i$ are called (anti) self-dual 2-forms. Any 2-form
can be decomposed into self-dual and anti self-dual parts 
\[
{\cal P}={\cal P}^{+}+{\cal P}^{-},\qquad \mbox{where}\quad *{\cal P}^{\pm
}=\pm i{\cal P}\,. 
\]
The Hodge star relates the hook and wedge operations by 
\begin{equation}  \label{hookwedge}
X%
\hbox{\vrule  height 0ex     width 1ex      depth 0.11ex
                 \vrule  height 1.6ex   width 0.11ex   depth 0.11ex
                 \vrule  height 0ex     width 0.5ex    depth 0ex   } *{\cal P%
} = *\left({\cal P}\wedge X^\flat\right)\,.
\end{equation}
The 2-form commutator is given by 
\begin{equation}
\lbrack {\cal P},{\cal Q}]=-2X_a%
\hbox{\vrule  height 0ex     width 1ex      depth 0.11ex
                 \vrule  height 1.6ex   width 0.11ex   depth 0.11ex
                 \vrule  height 0ex     width 0.5ex    depth 0ex   } {\cal P}%
\wedge X^a 
\hbox{\vrule  height 0ex     width 1ex      depth 0.11ex
                 \vrule  height 1.6ex   width 0.11ex   depth 0.11ex
                 \vrule  height 0ex     width 0.5ex    depth 0ex   } {\cal Q}
\label{2comm}
\end{equation}
for 2-forms ${\cal P}$ and ${\cal Q}$. The Lie algebra of 2-forms under
commutation is the Lie algebra of the Lorentz group.

It is often useful to work with a null coframe (basis for 1-forms) $\{l,n,m,%
\bar{m}\}$ dual to a Newman-Penrose tetrad, that is, one for which all inner
products vanish except 
\begin{equation}
l\cdot n=-m\cdot \bar{m}=1\,.  \label{np_dots}
\end{equation}
From this we can construct a basis for the anti self-dual 2-forms: 
\begin{equation}
{\cal U}=-n\wedge \bar{m},\quad {\cal M}=l\wedge n-m\wedge \bar{m},\quad 
{\cal V}=l\wedge m  \label{bivect_basis}
\end{equation}
with the property that all inner products vanish except 
\begin{equation}
{\cal U}\cdot {\cal V}=1\qquad \mbox{and}\qquad {\cal M}\cdot {\cal M}=-2\,.
\label{bivec_dots}
\end{equation}
In this basis, the 2-form commutator can be calculated from 
\begin{equation}
\lbrack {\cal M},{\cal U}]=-4{\cal U}\,,\quad [{\cal M},{\cal V}]=4{\cal V}%
\quad \mbox{and}\quad [{\cal U},{\cal V}]=-{\cal M}\,.  \label{bivec_comm}
\end{equation}
The null basis elements ${\cal U}$ and ${\cal V}$ each have one
two-dimensional eigenspace, with corresponding zero eigenvalue, spanned by $%
\{n^{\sharp },\bar{m}^{\sharp }\}$ and $\{l^{\sharp },m^{\sharp }\}$
respectively. These are also the eigenspaces of ${\cal M}$ for which they
have eigenvalues $+1$ and $-1$. Note that choosing ${\cal M}$ determines $%
{\cal U}$ and ${\cal V}$ up to their relative scaling or interchange.

We denote the torsion-free metric compatible covariant derivative of a
2-form ${\cal Q}$ with respect to a vector field $Z$ by $\nabla _Z{\cal Q}$.
In terms of this, the exterior derivative $d$ and co-derivative $\delta =*d*$
can be expressed: 
\begin{eqnarray*}
d &\equiv &e^a\wedge \nabla _{X_a}\,, \\
\delta &\equiv &-X^a%
\hbox{\vrule  height 0ex     width 1ex      depth 0.11ex
                 \vrule  height 1.6ex   width 0.11ex   depth 0.11ex
                 \vrule  height 0ex     width 0.5ex    depth 0ex   } \nabla
_{X_a}\,.
\end{eqnarray*}

\section{THE PETROV CLASSIFICATION}

\label{App:Petrov}

In a vacuum background, the Riemann curvature tensor ${\bf R}$ is equal to
the Weyl conformal curvature tensor ${\bf C}$. The symmetries of these
tensors allow them to be written as the sum of terms made of symmetric
tensor products of 2-forms (i.e. terms like ${\cal P}\otimes {\cal Q}+{\cal P%
}\otimes {\cal Q}$). So, both can be considered as self-adjoint maps on
2-forms; if $C_{abcd}$ are components of ${\bf C}$ and ${\cal P}_{ab}$ the
components of a 2-form, then the definition 
\[
\left( {\bf C}{\cal P}\right) _{ab}={\frac 12}C_{abcd}{\cal P}^{cd} 
\]
gives the components of the 2-form ${\bf C}{\cal P}$. As a map on 2-forms,
the conformal tensor preserves the eigenspaces of $*$ and so may decomposed
into a part made from self-dual 2-forms alone and a part made from anti
self-dual 2-forms. That is, we can write 
\[
{\bf C}={\bf C}^{(+)}+{\bf C}^{(-)}\,, 
\]
where ${\bf C}^{(\pm )}{\cal Q}^{\mp }=0$. Note that since the conformal
tensor is real, ${\bf C}^{(-)}$ is the complex conjugate of ${\bf C}^{(+)}$,
and so it is sufficient to classify only one of these.

The action of ${\bf C}^{(-)}$ on the Newman-Penrose 2-form basis described
in Appendix \ref{App:forms} is the same as the action of ${\bf C}$ on this
basis and can be written as 
\[
{\bf C}^{(-)}\left[ 
\begin{array}{c}
{\cal U} \\ 
{\cal M} \\ 
{\cal V}
\end{array}
\right] =\left[ 
\begin{array}{ccc}
-\Psi _2 & \Psi _3 & -\Psi _4 \\ 
-2\Psi _1 & 2\Psi _2 & -2\Psi _3 \\ 
-\Psi _0 & \Psi _1 & -\Psi _2
\end{array}
\right] \left[ 
\begin{array}{c}
{\cal U} \\ 
{\cal M} \\ 
{\cal V}
\end{array}
\right] 
\]
Note that the matrix of this transformation is trace-free and the mapping is
self-adjoint (that is, ${\cal Q}\cdot {\bf C}{\cal P}={\bf C}{\cal Q}\cdot 
{\cal P}$).

The Petrov classification is a classification of this mapping. The spacetime
is known as algebraically general when there are three distinct eigenvalues,
and algebraically special otherwise. Two special cases of interest here are
that of type $D$ and $N$, for which a basis can be chosen so that the matrix
above takes the forms, 
\[
\left[ 
\begin{array}{ccc}
-\Psi _2 & 0 & 0 \\ 
0 & 2\Psi _2 & 0 \\ 
0 & 0 & -\Psi _2
\end{array}
\right] \qquad \mbox{and}\qquad \left[ 
\begin{array}{ccc}
0 & 0 & 0 \\ 
0 & 0 & 0 \\ 
-\Psi _0 & \ \ 0\ \  & \ \ 0\ \ 
\end{array}
\right] 
\]
respectively.

The real null direction of a null anti self-dual 2-form ${\cal Q}$ is said
to be a {\em principal null direction} (PND) of the conformal tensor if $%
{\cal Q}\cdot {\bf C}{\cal Q}=0$. We will call such a ${\cal Q}$, a {\em %
principal null} (PN) 2-form. There can be at most four independent PNDs and
their number and ``multiplicities'' provide another description of the
Petrov types \cite{pen&rin}. The multiplicities can be determined in the
present formulation by the following (with ${\cal P}$ an anti self-dual
2-form):

\begin{center}
\begin{tabular}{cccc}
{\bf multiplicity} &  & {\bf equivalent conditions} &  \\ 
1 & ${\cal Q}\cdot {\bf C}{\cal Q}=0$ &  & $\Psi _4=0$ \\ 
2 & $[{\cal Q},{\bf C}{\cal Q}]=0$ & ${\bf C}{\cal Q}\propto {\cal Q}$ & $%
\Psi _3=\Psi _4=0$ \\ 
3 & ${\cal Q}\cdot {\bf C}{\cal P}=0\quad \forall {\cal P}$ & ${\bf C}{\cal Q%
}=0$ & $\Psi _2=\Psi _3=\Psi _4=0$ \\ 
4 & \quad $[{\cal Q},{\bf C}{\cal P}]=0\quad \forall {\cal P}$ \quad & \quad 
${\bf C}{\cal P}\propto {\cal Q}\quad \forall {\cal P}$ \quad & \quad $%
\Psi_1=\Psi _2=\Psi _3=\Psi _4=0$%
\end{tabular}
\end{center}

\section{CKY 2-FORMS AND SHEAR-FREE CONGRUENCES}

\label{App:shearfree}

Defining the shear of a null geodesic vector field requires the choice of a
``screen space'', and so is not an intrinsic property of the vector field.
However, if the shear vanishes for one choice of screen space, then it does
for all and hence the notion of a shear-free null vector field is well
defined. For definitions and discussion of optical scalars see \cite{pen&rin}%
.

Robinson \cite{rob} showed that the real null eigenvector $l$ of a (anti)
self-dual null 2-form $\phi $ is geodesic and shear-free if and only if $%
\phi $ is proportional to a source-free Maxwell field, that is $d\phi =0$.
Note that the eigenspace such a 2-form is two-dimensional, isotropic and
integrable. So we can use this fact or the Frobenius integrability
condition, that $d\phi =\alpha \wedge \phi $ for some $\alpha $, for the
vanishing of the shear of $l$. It is convenient here to use these results
interchangeably as our criterion for a shear-free null geodesic.

Note that a shear-free null geodesic is a PND of the conformal tensor.

\subsection{Null CKY 2-forms}

Now, suppose that ${\cal Q}$ is a null anti self-dual CKY 2-form. Since the
right hand side of CKY 2-form Eq.(\ref{formcky2}) is simply the anti
self-dual part of $-2Z^{\flat }\wedge \delta {\cal Q}$, we have that 
\begin{eqnarray*}
0={\cal Q}\cdot 3\nabla _Z{\cal Q} &=&-2(Z^{\flat }\wedge \delta {\cal Q}%
)\cdot {\cal Q} \\
&=&2Z%
\hbox{\vrule  height 0ex     width 1ex      depth 0.11ex
                 \vrule  height 1.6ex   width 0.11ex   depth 0.11ex
                 \vrule  height 0ex     width 0.5ex    depth 0ex   } (\delta 
{\cal Q})^{\sharp }%
\hbox{\vrule  height 0ex     width 1ex      depth 0.11ex
                 \vrule  height 1.6ex   width 0.11ex   depth 0.11ex
                 \vrule  height 0ex     width 0.5ex    depth 0ex   } {\cal Q}%
\,.
\end{eqnarray*}
Hence we can find an $\alpha $ such that $\delta {\cal Q} =\alpha ^{\sharp }%
\hbox{\vrule  height 0ex     width 1ex      depth 0.11ex
                 \vrule  height 1.6ex   width 0.11ex   depth 0.11ex
                 \vrule  height 0ex     width 0.5ex    depth 0ex   } {\cal Q}
$ or equivalently $d{\cal Q}=-\alpha \wedge {\cal Q}$. So the real null
eigenvector of ${\cal Q}$ is shear-free.

\subsection{Non-null CKY 2-forms}

We wish to show that the eigenspaces of a non-null CKY 2-form ${\cal Q}$ are
integrable and hence contain a shear-free null geodesic vector field. That
is, we want to show that if $X$ and $Y$ are elements of the same eigenspace
of ${\cal Q}$ with eigenvalue $\lambda $ ($X%
\hbox{\vrule  height 0ex     width 1ex      depth 0.11ex
                 \vrule  height 1.6ex   width 0.11ex   depth 0.11ex
                 \vrule  height 0ex     width 0.5ex    depth 0ex   }{\cal Q}%
=\lambda X^{\flat }$ and $Y%
\hbox{\vrule  height 0ex     width 1ex      depth 0.11ex
                 \vrule  height 1.6ex   width 0.11ex   depth 0.11ex
                 \vrule  height 0ex     width 0.5ex    depth 0ex   }{\cal Q}%
=\lambda Y^{\flat }$), then so is $[X,Y]$. Since $[X,Y]=\nabla _XY-\nabla
_YX $, we will show that $\nabla _XY%
\hbox{\vrule  height 0ex     width 1ex      depth 0.11ex
                 \vrule  height 1.6ex   width 0.11ex   depth 0.11ex
                 \vrule  height 0ex     width 0.5ex    depth 0ex   }{\cal Q}%
=\lambda \nabla _XY^{\flat }$. Note that this eigenspace is isotropic, that
is $g(X,Y)=0$.

Since the map $\alpha \mapsto \alpha ^{\sharp }%
\hbox{\vrule  height 0ex     width 1ex      depth 0.11ex
                 \vrule  height 1.6ex   width 0.11ex   depth 0.11ex
                 \vrule  height 0ex     width 0.5ex    depth 0ex   }{\cal Q}$
is of maximal rank for non-null ${\cal Q}$, it can always be inverted and a
1-form $\alpha $ found such that $\delta {\cal Q}=-\alpha ^{\sharp }%
\hbox{\vrule  height 0ex     width 1ex      depth 0.11ex
                 \vrule  height 1.6ex   width 0.11ex   depth 0.11ex
                 \vrule  height 0ex     width 0.5ex    depth 0ex   }{\cal Q}$
and $d{\cal Q}=\alpha \wedge {\cal Q}$. Using these expressions for $\delta 
{\cal Q}$ and $d{\cal Q}$, and the CKY 2-form Eq.(\ref{formcky2}), we have 
\begin{eqnarray*}
\nabla _XY%
\hbox{\vrule  height 0ex     width 1ex      depth 0.11ex
                 \vrule  height 1.6ex   width 0.11ex   depth 0.11ex
                 \vrule  height 0ex     width 0.5ex    depth 0ex   }{\cal Q}
&=&\nabla _X(Y%
\hbox{\vrule  height 0ex     width 1ex      depth 0.11ex
                 \vrule  height 1.6ex   width 0.11ex   depth 0.11ex
                 \vrule  height 0ex     width 0.5ex    depth 0ex   }{\cal Q}%
)-Y%
\hbox{\vrule  height 0ex     width 1ex      depth 0.11ex
                 \vrule  height 1.6ex   width 0.11ex   depth 0.11ex
                 \vrule  height 0ex     width 0.5ex    depth 0ex   }\nabla _X%
{\cal Q} \\
&=&\lambda \nabla _XY^{\flat }+X\lambda Y^{\flat }-{\frac 13}\lambda \alpha
(X)Y^{\flat }.
\end{eqnarray*}
Rearranging and writing the vector equation dual to this shows that 
\begin{equation}
\left( \nabla _XY%
\hbox{\vrule  height 0ex     width 1ex      depth 0.11ex
                 \vrule  height 1.6ex   width 0.11ex   depth 0.11ex
                 \vrule  height 0ex     width 0.5ex    depth 0ex   }{\cal Q}%
\right) ^{\sharp }-\lambda \nabla _XY=\left( X\lambda -{\frac 13}\lambda
\alpha (X)\right) Y\,.  \label{eig1}
\end{equation}
Note that the right hand side is a multiple of $Y$ and hence an eigenvector
of ${\cal Q}$ with eigenvalue $\lambda $. However, upon contracting the left
hand side with ${\cal Q}$, we find that it is an element of the other
eigenspace, having eigenvalue $-\lambda $. Hence we must conclude that 
\begin{equation}
\nabla _XY%
\hbox{\vrule  height 0ex     width 1ex      depth 0.11ex
                 \vrule  height 1.6ex   width 0.11ex   depth 0.11ex
                 \vrule  height 0ex     width 0.5ex    depth 0ex   }{\cal Q}%
-\lambda \nabla _XY^{\flat }=0\,,  \label{eig3}
\end{equation}
and we have the required result.

Since each eigenspace of ${\cal Q}$ is integrable they each give rise to a
null self-dual 2-form proportional to a Maxwell field, and hence the real
eigenvectors of ${\cal Q}$ are shear-free.

\section{INTEGRABILITY OF CKY 2-FORMS}

\label{App:CKYint}

Apart from conformally flat spacetimes, CKY 2-forms can only exist in
spacetimes of Petrov type $D$ or $N$. To understand this it is sufficient to
consider only CKY tensors of definite Hodge-duality, for which we give an
integrability condition. For an anti self-dual CKY 2-form ${\cal Q}$, 
\begin{equation}
\lbrack {\cal Q},C{\cal P}]={\frac 12}[{\cal P},C{\cal Q}]\,,\qquad %
\mbox{$\forall$ 2-forms ${\cal P}$}.  \label{cqcomm}
\end{equation}
If we let ${\cal P}={\cal Q}$, it follows that 
\[
\lbrack {\bf C}{\cal Q},{\cal Q}]=0.
\]
Then, from the commutator algebra of anti self-dual 2-forms Eq.(\ref
{bivec_comm}), it can be deduced that $C{\cal Q}$ must be proportional to $%
{\cal Q}$, i.e. 
\begin{equation}
{\bf C}{\cal Q}=\mu {\cal Q},  \label{cqmu}
\end{equation}
where $\mu $ is a scalar. From this, we can deduce the Petrov type as
described in Appendix \ref{App:Petrov}.

\subsection{Null CKY 2-forms}

When ${\cal Q}$ is null this implies that the real null eigenvector of $%
{\cal Q}$ is a repeated principal null direction. However, if we write out
Eq.(\ref{cqcomm}) in an anti self-dual 2-form basis chosen so that ${\cal U}=%
{\cal Q}$ and ${\cal V}\propto {\cal P}$, we find that $\mu =-\Psi _2=0 $.
Not only does this immediately tell us that $C{\cal Q}=0$, but upon
substitution into Eq.(\ref{cqcomm}) we have that $[{\cal Q},C{\cal P}]=0$
for all anti self-dual 2-forms ${\cal P}$. Hence the real null direction
defined by ${\cal Q}$ is a four-fold PND and the spacetime is of Petrov type 
$N$.

\subsection{Non-null CKY 2-forms}

When ${\cal Q}$ is non-null, we concluded in Appendix \ref{App:shearfree}
that the real null eigenvectors of ${\cal Q}$ are shear-free. If we align
our anti self-dual 2-form basis so that ${\cal M}\propto {\cal Q}$ then $%
{\cal U}$ and ${\cal V}$ have shear-free eigenvectors and hence are PN
2-forms. From this we conclude that $\Psi _0=\Psi _4=0$. The integrability
condition Eq.(\ref{cqmu}) immediately requires that $\Psi _1$ and $\Psi _3$
vanish and hence the spacetime is of Petrov type $D$.

This reasoning made no use of Ricci-flatness wherein the Goldberg-Sachs
theorem \cite{josh&ray} would imply the same result.

\end{document}